\documentclass{llncs}
\usepackage[dvips]{graphicx}
\begin{document}
%
%
%
\title{A distributed editing environment for XML documents}

\author{Claude Pasquier \and Laurent Th\'ery}
\authorrunning{Claude Pasquier and Laurent Th\'ery}
\institute{Inria Sophia Antipolis,\\ 2004, route des Lucioles - BP 93,\\
06902 Sophia Antipolis, France}

\maketitle              

\begin{abstract}
XML is based on two essential aspects: the modelization of data in a
tree like structure and the separation between the information itself
and the way it is displayed. XML structures are easily serializable.
The separation between an abstract representation and one or
several views on it allows the elaboration of specialized interfaces
to visualize or modify data. A lot of developments were made to
interact with XML data but the use of these applications over the
Internet is just starting.\\
This paper presents a prototype of a distributed editing environment
over the Internet. The key point of our system is the way user interactions 
are handled. Selections and modifications made by a
user are not directly reflected on the concrete view, they are
serialized in XML and transmitted to a server which applies them to
the document and broadcasts updates to the views.\\
This organization has several advantages. XML documents coding
selection and modification operations are usually smaller than the
edited document and can be directly processed with a transformation
engine which can adapt them to different representations. In addition,
several selections or modifications can be combined into an unique XML
document. This allows one to update multiple views with different
frequencies and fits the requirement of an asynchronous communication
mode like HTTP.
\end{abstract}
\section{Introduction}
XML is the central point of the far-reaching process of standardization
that is going to alter the way information is handled. The deployment
of XML related technologies will have a large impact on operations
like structured editing, storage, information exchange, data
transformation, querying, and rendering.

Two essential aspects of XML have been inherited from SGML~\cite{sgml}: the
modelization of data in a tree like structure and the separation
between information itself and the way it is displayed. An XML
document is typically a serialization of a tree with every node and
leaf tagged. It is a convenient user-readable and platform-independent
representation, well-suited for transmission over a network. The
separation between an abstract representation and one or several views
on it allows the elaboration of specialized interfaces to visualize or
modify data.

A lot of developments were made to interact with XML data but the use
of these applications over the Internet is still under development. The
World Wide Web Consortium proposes several standards to visualize XML
documents (CSS~\cite{css}, XSL~\cite{xsl}) and to transform them (XSLT
\cite{xslt}). However, these
standards are mainly defined to display documents, not to interact
with them. If we consider real e-commerce applications, displaying
and modifying data will be needed. For example, a vendor may want to
update via Internet (using a portable computer or a mobile phone) the
database of its company with new information relative to a visited
client.

On new XML browsers, some interaction with a displayed document can
nevertheless be realized. Documents are internally modelized as DOM
trees and can be manipulated with methods that access public DOM
APIs. The problem with this solution is that only the concrete view of
a document is edited, not the data itself. 
This paper presents a prototype of a distributed editing environment
over the Internet. Its architecture is based on the
software development environment Centaur~\cite{bor88} and ideas presented
in~\cite{cle90} and~\cite{dery}.
The key point of our system is the way user interactions are
handled. Selections and modifications made by a user are not directly
reflected on the concrete view. They are serialized in XML and
transmitted to a server which applies them to the document and
broadcasts updates to the views. Bidirectional correspondences
between a source and a result tree are expressed in the declarative
language XPPML. It is used to transform serialized user-interactions
made on the source structure to equivalent operations applicable to
the result structure.

\section{Interacting with XML documents}
Editing an XML document over an asynchronous, unstable and rather slow
protocol like HTTP implies minimizing the amount of information
exchanged. In particular when editing large documents, it is not a
very good strategy to retransmit every time the modified document.
User interactions are encoded as XML documents. Data transmitted 
on the network represent actions. The first set of actions declares
external operations on the structure (selection,
modification, redisplay, etc). The second set concerns updates to be
applied on the structure (selection of a given subtree, deletion of a
node, etc.). In our prototype, we have defined two actions, selection
and modification, that are both based on an unambiguous way to
identify a node on a tree.

We take as an example the simple mathematical expression {\em '2*(8/2+5)'}
which will be used throughout this paper. It is expressed by the
following XML document called {\em 'exp'}:
\begin{verbatim}
<?XML version='1.0'?>
<exp>
  <mult>
    <int value='2'/>
    <plus>
      <div>
        <int value='8'/>
        <int value='2'/>
      </div>
      <int value='5'/>
    </plus>
  </mult>
</exp>
\end{verbatim}

\subsection{Paths}
One can identify a node in the tree with an expression
specifying a path to this node. This can be expressed with the XPath
\cite{xpath} standard by starting from the document's root and specifying
either the name of all the encountered child nodes or their relative
positions. For example, the node {\em '8/2+5'} can be equally designated
by the following XPath expressions:
\begin{verbatim}
/exp/mult/plus
/*[1]/*[1]/*[2]
\end{verbatim}
Note that identifying subtrees by a list of node names is ambiguous, since
two different subtrees could have the same path. For this reason, we prefer the second 
solution. In order to transmit paths, we propose to 
represent a XPath location which uses relative positions 
by an ipath (IntegerPath) XML element. The location 
of the node {\em '8/2+5'} is thus expressed by:

\begin{verbatim}
<ipath>
  <move num='1'/>
  <move num='2'/>
</ipath>
\end{verbatim}
The starting point of the path is the root node (called
documentElement in DOM) and not ``the parent of the document element'' 
as it is specified in XPath. Then, the empty element {\tt <ipath/>}
identifies the root of the document, {\tt <ipath>
<move num='1'/></ipath>} its first child.
\subsection{Selection path}
We use selection path to designate different elements of a document.
These elements may belong to different selections. Selections
are represented by symbolic names. A {\em spath} element is defined 
by the following DTD:
\begin{verbatim}
<!ELEMENT spath  (select*, (move, spath)*)>
<!ELEMENT select EMPTY>
<!ATTLIST select name NMTOKEN #REQUIRED>
<!ELEMENT move   EMPTY>
<!ATTLIST move   num  NMTOKEN #REQUIRED>
\end{verbatim}
A {\em spath} element is composed of two sets of elements. The first
one is used to declare and name the selections on the current node. The
second one is composed of pairs of {\em move} and {\em spath}
elements. Each {\em spath} represents the selection path corresponding 
to the sub-element denoted by the {\em move}.

For example, a selection called {\em 'selA'} of the node {\em '5'} and a selection
{\em 'selB'} of the node {\em '8/2+5'} is expressed by the following spath element:
\begin{verbatim}
<spath>
  <move num='1'/>
  <spath>
    <move num='2'/>
    <spath>
      <select name='selB'/>
      <move num='2'/>
      <spath>
        <select name='selA'/>
      </spath>
    </spath>
  </spath>
</spath>
\end{verbatim}
Selection paths are general enough to be used for different purposes.
On the client side it can be used to represent extension to an existing 
selection. On the server side it can be used to represent all the existing 
selections.

\subsection{Modification path}
A modification path memorizes modifications done on a document.
The following DTD describes the structure of a {\em mpath} element:
\begin{verbatim}
<!ELEMENT mpath  (element, (move, mpath)*)>
<!ATTLIST mpath type (move|delete|insert|change) move>
<!ELEMENT element (%targetElt;)>
<!ELEMENT move   EMPTY>
<!ATTLIST move   num  NMTOKEN #REQUIRED>
\end{verbatim}
A {\em mpath} has an attribute {\em type} which indicates the kind of
modification to perform (deletion, insertion or
replacement). By default, if no attribute is given, the type is {\em move}.
Values that are changed or inserted are specified under the tag {\em element}.

As an example, one can evaluate the node {\em '8/2'} and replace it
with its integer value. This is expressed by the following {\em mpath} 
element:
\begin{verbatim}
<mpath>
  <move num='1'/>
  <mpath>
    <move num='2'/>
    <mpath>
      <move num='1'/>
      <mpath type='change'>
        <element>
          <int value='4'/>
        </element>
      </mpath>
    </mpath>
  </mpath>
</mpath>
\end{verbatim}
A modification path can be applied to a document to perform the memorized 
operations. Note that this application can be delayed. Two {\em mpath} elements 
can be combined in a new modification path which amalgamates the successive
operations. For example, the previous modification path and the one that
corresponds to the deletion of the node identified by the selection 
{\em 'selA'} gives the following combined path:
\begin{verbatim}
<mpath>
  <move num='1'/>
  <mpath>
    <move num='2'/>
    <mpath>
      <move num='1'/>
      <mpath type='change'>
        <element>
          <int value='4'/>
        </element>
      </mpath>
    <move num='2'/>
    <mpath type='delete'/>
  </mpath>
</mpath>
\end{verbatim}
When applied to a document, this {\em mpath} will replace the element {\em '8/2'} 
with {\em '4'} and will remove the element {\em '5'}. The same {\em mpath} can also 
be applied to the current selection to compute a new selection path
compatible with the updated structure of the document.

Modification paths are generic enough to represent any kind of 
modification. They represent a convenient way to interact with
large documents over a network since only the new elements of the
document need to be transmitted.
\section{Communication with a user interface}
Except for XML editors that let the user directly write the tags of XML
documents, standard editing environments propose an interface between
the user and the logical structure. Usually, selections and
modifications are done on a concrete view of the data. User
interactions are transformed into actions on the logical structure and
the layout of the  view is recomputed.

Our concepts of selection and modification paths can be applied to
modelize the communication between a concrete view and an abstract
representation. Representing actions by valid XML documents makes it
possible to use standard tools to manipulate them. Let's take the
example of an XML browser in which the displayed document is obtained 
by transforming an initial document. All modifications made on the source 
document can be reflected to the view by transmitting only a modification 
path. 

Modification paths on the logical tree and the concrete tree can largely 
differ. However, one can be generated from the other by using a 
transformation process similar to the one used to get the initial
document. This means that the transformation tool should 
be capable not only of processing the initial document but also of
transforming selection and modification paths from one structure to 
another.

Standard transformation languages, like DSSSL~\cite{dsssl} or XSLT
\cite{xslt} are well-suited for  processing  XML
documents. However, current implementations of these languages are
based on a batch process that transforms the whole source document into 
a target one. Once this is done, no link is kept between the two 
structures. In addition,  there is no available implementation capable
of doing a reverse transformation (from a target structure back to the 
source one). Adding this dynamic capability to engines based on DSSSL or
XSLT standards seems difficult because of the expressiveness of
these languages. In our project, we have developed a transformation engine 
based on XPPML (Xml Pretty-Printing Meta Language) that is
strictly less powerful than XSLT but  satisfies all our dynamic requirements. 
It is a modified version of the transformation engine 
developed for the A\"ioli system~\cite{aioli}.
\section{XPPML}
XPPML is an XML extension for the pretty-printing meta language PPML
\cite{ppml} defined in the Centaur system. An XPPML specification is a
collection of unparsing rules associated with abstract syntax
patterns. The concepts of XPPML are very close to those found in
XSLT. Basically, it is a language for transforming XML documents into
other representations. A transformation in the XPPML language is
expressed as a well-formed XML document. It describes rules for
converting a source tree into a result one by associating patterns
with templates. The formatting machine generates a result structure by
traversing the source tree and looking for a pattern that matches each
node. When a match is found, the corresponding template is
instantiated to create parts of the result tree. Features of XPPML
include contextual formatting, conditional layout over external
boolean functions and inclusion of user-defined external functions.

The formatting machine makes use of the XPPML rules to generate the path on
the result tree corresponding to a given path on the source one but is
also able to retrieve the position on a source structure corresponding 
to a position on a result one.

A typical XPPML document has the following structure:
\begin{verbatim}
<x:prettyPrinter ppName="..." langName="..."
         xmlns:x='http://www-sop.inria.fr/lemme/xppml/1.0'>
  <x:extension name="..."/>
  <x:import package="..."/>
  <x:rule>
    ...
  </x:rule>
</x:prettyPrinter>
\end{verbatim}
An XPPML specification is fully identified by both the name of the
pretty printer and the name of the language to which it applies. This
allows one to refer and retrieve, for example, the standard pretty
printer for the Java language or the pretty printer Y for the language
Z. These two identifiers are stored in the attributes {\em 'ppName'} and
{\em 'langName'} of the element {\em 'prettyPrinter'}.

Modularity of pretty-printing declarations is realized through one or
several {\em 'extension'} elements where the names of other XPPML
definitions to be included are specified. The {\em 'import'} element is used
to declare the package where external Java functions should be
searched.

The core of the XPPML declaration is defined by a list of {\em 'rule'}
elements composed of a {\em 'pattern'} element and a layout. A pattern is
defined by a {\em 'template'} element and zero or more additional
constraints. The {\em 'match'} attribute in the {\em 'template'} element is a
pattern that identifies the source node to which the rule applies. For
example, the following XPPML specification defines the layout of our
{\em 'exp'} document:
\begin{verbatim}
<x:prettyPrinter ppName='std' langName='exp'
         xmlns='http://www-sop.inria.fr/lemme/figue/1.0'
         xmlns:x='http://www-sop.inria.fr/lemme/xppml/1.0'>
  <x:rule>
    <x:pattern><x:template match='mult(*x,*y)'/></x:pattern>
    <h>
     <x:variable name='x'/>
     <atom value='*'/>
     <x:variable name='y'/>
    </h>
  </x:rule>
  <x:rule>
    <x:pattern><x:template match='mult/plus(*x,*y)'/></x:pattern>
    <h>
      <atom value='('/>
      <x:variable name='x'/>
      <atom value='+'/>
      <x:variable name='y'/>
      <atom value=')'/>
    </h>
  </x:rule>
  <x:rule>
    <x:pattern><x:template match='plus(*x,*y)'/></x:pattern>
    <h>
      <x:variable name='x'/>
      <atom value='+'/>
     <x:variable name='y'/>
    </h>
  </x:rule>
  <x:rule>
    <x:pattern><x:template match='x=int'/></x:pattern>
    <x:extFun name='identitypp'>
      <x:arg value='x' type='var'/>
    </x:extFun>
  </x:rule>
</x:prettyPrinter>
\end{verbatim}
XML namespaces are used to distinguish between XPPML constructs and
elements corresponding to the output structure. This example uses
Figue's syntax~\cite{figue} for the layout but other output format like HTML or
xsl:fo can equally be used. Figue is an incremental bi-directional
layout engine that handles a limited set of combinators like {\em 'h'} to
specify an horizontal layout, {\em 'v'} for a vertical one, {\em 'atom'} for a
terminal box and some other specialized mathematical constructs.

The first  rule matches a {\em 'mult'} node composed of exactly two
children. Its layout part specifies that an horizontal box has to be
created (use of Figue's horizontal combinator {\em 'h'}) and filled with the
concrete tree obtained by a recursive call of the formatting machine
on the node's first child, a concrete leaf with the textual value {\em 
'*'} and the layout of the second child.

Contextual pattern are specified in the left-hand side of a rule, as in
the second rule which matches a  {\em 'plus'} node that have a {\em 'mult'} node
as parent. It is also possible to define rules identified by a context
name. When a contextual pretty-printing is required, a context name is
added to the recursive call.

In the fourth rule, an external Java method called {\em 'identitypp'} is
used to display the attribute {\em 'value'} of the node representing an
integer. External functions can also be used with conditional layout
constructs defined with the XPPML instructions {\em 'if'} and {\em 'case'}.

Editing and rendering are handled by Figue. During the editing
process, Figue communicates to the formatter the location of concrete
subtrees which have been selected or modified. The formatter retrieves
the corresponding abstract trees, performs the necessary modifications
and returns back to the layout engine the selection or modification
paths applicable on the concrete tree. With this information, Figue
performs an incremental update on the view. 

A\"ioli can be used as a server which communicates with clients through
HTTP requests. Documents are stored on the server side where updating and 
formatting operations are performed. Information manipulated on the server 
side is represented by DOM objects. These are transformed to XML text format
and are transmitted to the client through HTTP(S). Standard HTTP protocol is
used; marshaling and unmarshaling are done with application methods.

In our current implementation, several clients can be connected to the
same server. Figue is one of the possible clients but standard web
browsers can also be used. New clients may join and watch the progress of
existing editing sessions but problems specific to the implementation
of a concurrent editing environment~\cite{shaul} have not yet been studied.

\section{A typical editing session}
If our example document is accessible at the url {\em 'docURL'} and our
server is located on {\em 'myserver'}, then a typical invocation of the
transformation engine is done by the following HTTP request:
\begin{verbatim}
http://myserver&doc=docURL&ppml=std&type=figue
\end{verbatim}
The arguments {\em 'ppml'} and {\em 'type'} correspond respectively to the name of
the prettyprinter to use and the type of syntax used to codify the
concrete tree. The client first declares a new selection called {\em current}:
\begin{verbatim}
clientMess #1: <setSelection type='Single' name='current'/>
\end{verbatim}
The server answers with a reference to the new selection. This information 
is needed because if several clients are accessing the same document, the 
name of the selection may be already used. In that case,  the server returns
a selection with a different name:
\begin{verbatim}
serverMess #1: <selection type="Single" name="current" />
\end{verbatim}
Then the client requests for a copy of the document using the appropriate
prettyprinting:
\begin{verbatim}
clientMess #2: <redraw />
\end{verbatim}
The server sends the corresponding concrete tree using the Figue
syntax:
\begin{verbatim}
serverMess #2: <redraw><H><Atom Value="2"/><Atom Value="*"/>
               <H><Atom Value="("/><H><Atom Value="8"/>
               <Atom Value="/"/><Atom Value="2"/></H>
               <Atom Value="+"/><Atom Value="5"/>
               <Atom Value=")"/></H></H></redraw>
\end{verbatim}
The client is now able to display the expression:
\vskip10pt
\centerline{
\includegraphics[width=10cm]{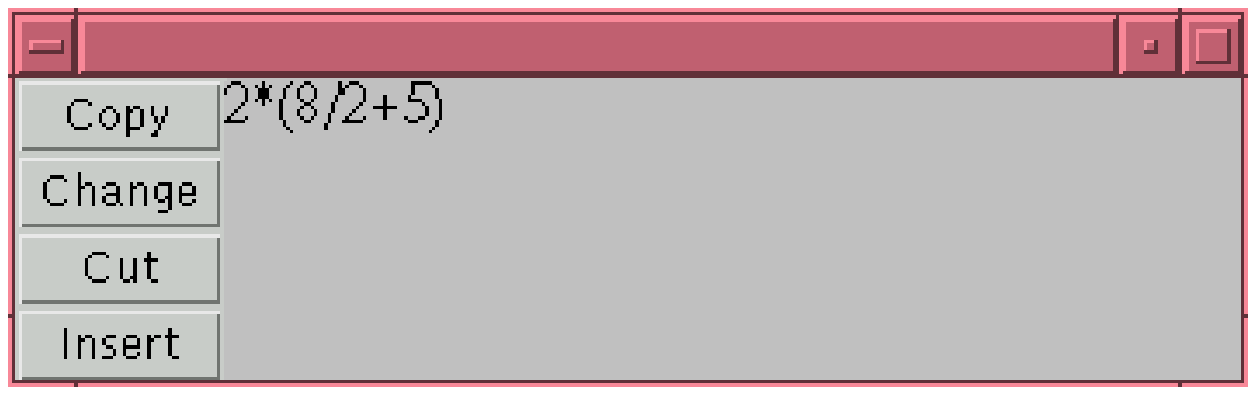}
}
\vskip10pt
\noindent
Let's suppose we want to edit the division. We first select the character 
{\em '/'} with the mouse:
\vskip10pt
\centerline{
\includegraphics[width=10cm]{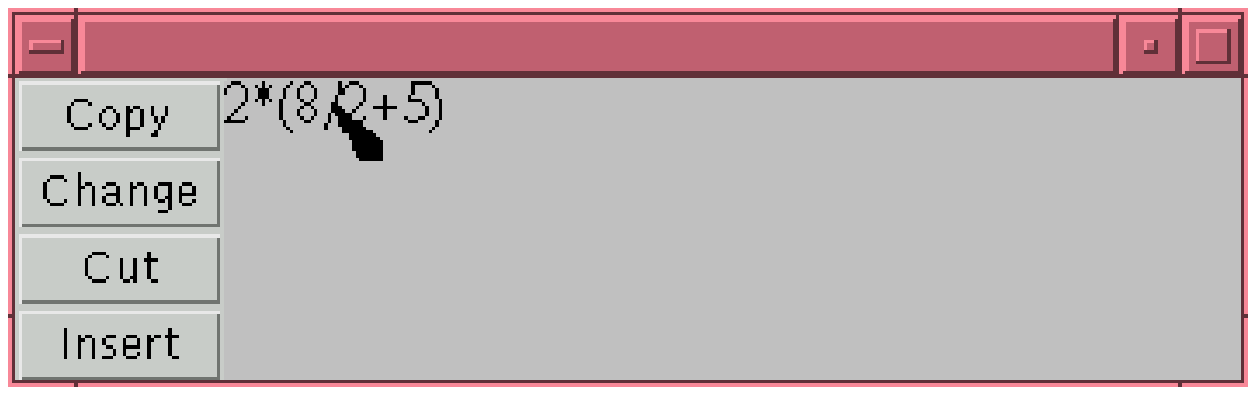}
}
\vskip10pt
\noindent
In order to update a selection, the client needs to send a message
to the server. It first computes the {\em ipath} corresponding to
the selected expression in the result tree and sends a message asking
for the {\em current} selection to be modified:
\begin{verbatim}
clientMess #3: <updateSelection selName='current'><ipath>
               <move num="3"/><move num="2"/><move num="2"/>
               </ipath></updateSelection>
\end{verbatim}
The server acknowledges this message:
\begin{verbatim}
serverMess #3: <done/>
\end{verbatim}
The server translates the selection made on the result tree into a selection
on the source tree. In our case, it is the tree {\tt <div><int value='8'/>
<int value='2'/></div>} that has generated the character {\em '/'}.

In order to get the value of the selection, the client requests the list of 
selection's changes that are memorized on the server side:
\begin{verbatim}
clientMess #4: <commit type="select"/>
\end{verbatim}
The server responds with a structure specifying that the {\em current} 
selection must  be placed on the second child of the third node, i.e. 
the node representing {\em '8/2'}.
\begin{verbatim}
serverMess #4: <commit type="select"><Change>
               <select name="current"/></Change><Path Rank="3">
               <Path Rank="2"><ExtendSelection Name="current" />
               </Path></Path></commit>
\end{verbatim}
This information can be used to highlight the current selection
in the editor:
\vskip10pt
\centerline{\includegraphics[width=10cm]{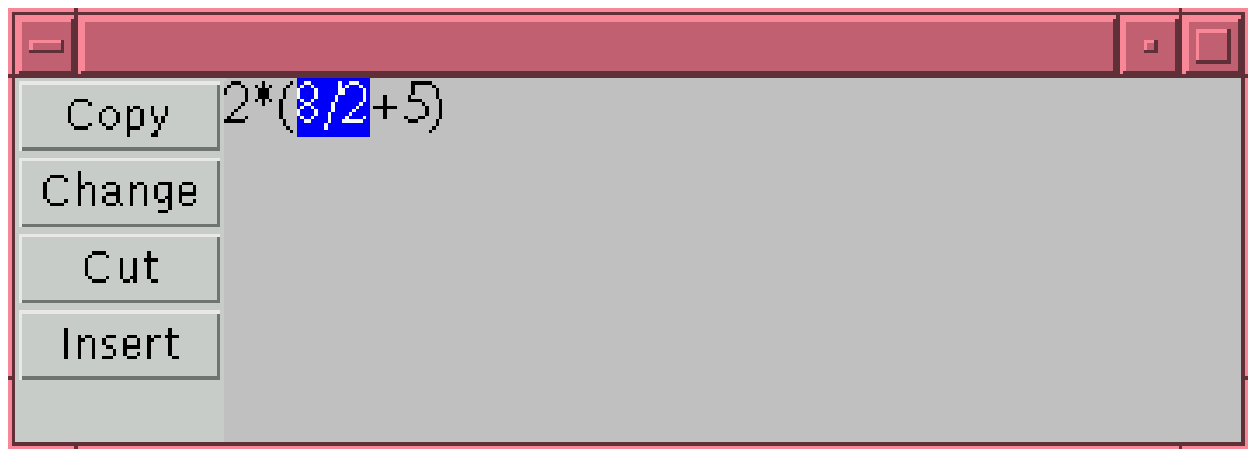}}
\vskip10pt
\noindent
Note that what we obtain is a structured selection: by selecting
the single character {\em '/'} we get the whole division expression.
Now that the division is selected, we can decide to evaluate it and
replace it with 4:
\begin{verbatim}
clientMess #5: <change selName='current'><int value="4"/></change>
\end{verbatim}
The modification is done on the abstract tree
stored on the server side and an acknowledgement is sent.
\begin{verbatim}
serverMess #5: <done/>
\end{verbatim}
To get the modification, the client just asks for the list of modifications
that have been done on the abstract structure since the last commit
request:
\begin{verbatim}
clientMess #6: <commit type="modif"/>
\end{verbatim}
The server returns the modifications to be applied on
the concrete tree:
\begin{verbatim}
serverMess #6: <commit type="modif"><mpath type="move">
               <move num="3"/><mpath type="move"><move num="2"/>
               <mpath type="change"><element><Atom Value="4"/>
               </element></mpath></mpath></mpath></commit>
\end{verbatim}
The client can then reflect these modifications on the editor:
\vskip10pt
\centerline{
\includegraphics[width=10cm]{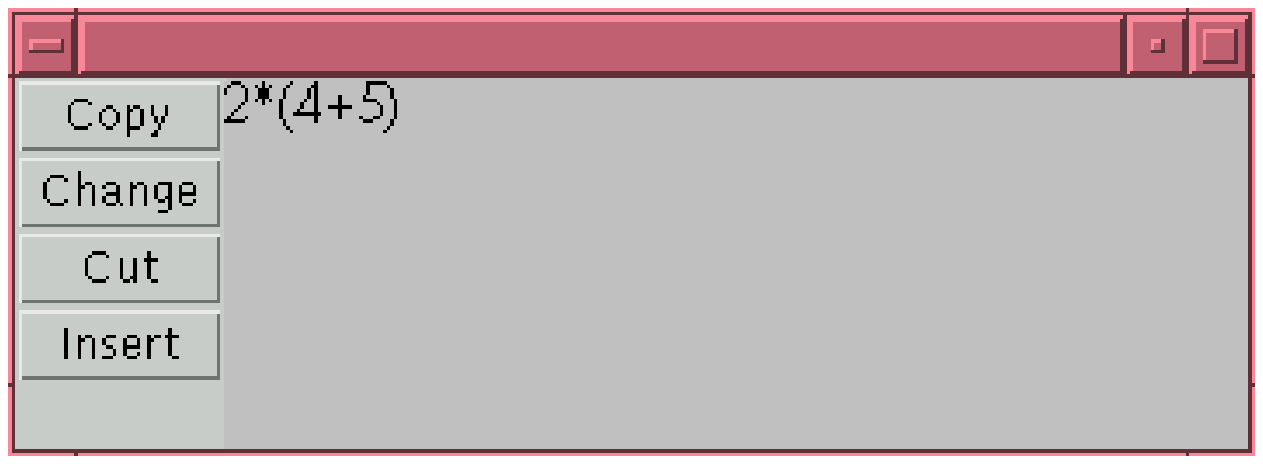}}
\vskip10pt
\noindent
The modification may have changed the values of some selections.
So, the client asks for their new values:
\begin{verbatim}
clientMess #7: <commit type="select"/>
\end{verbatim}
The modification has only replaced a selected tree; nothing
has changed:
\begin{verbatim}
serverMess #7: <commit type="select"><Change>
               <select name="current"/></Change><Path Rank="3">
               <Path Rank="2"><ExtendSelection Name="current"/>
               </Path></Path></commit>
\end{verbatim}
The client can update the editor with the selection:
\vskip10pt
\centerline{
\includegraphics[width=10cm]{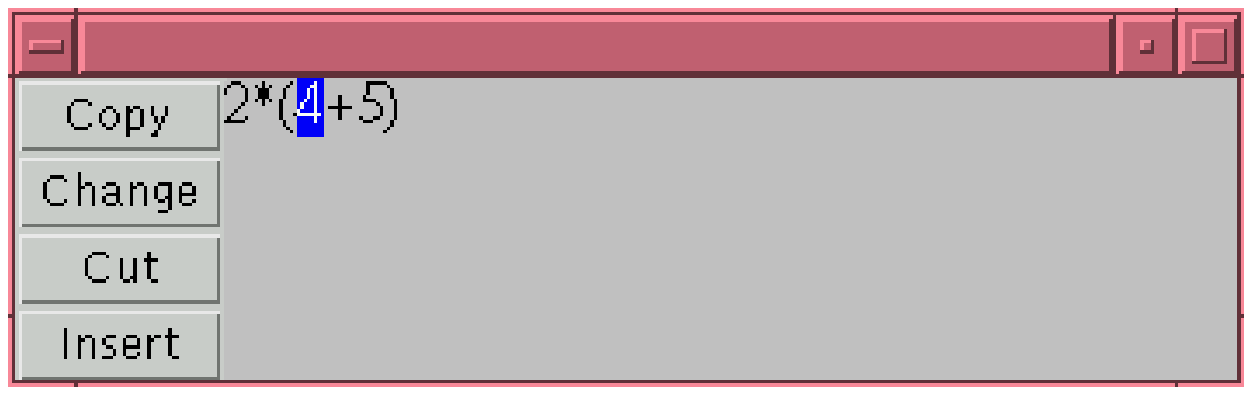}
}
\section{Conclusion}
The solution described in this paper has several advantages. XML
documents coding selection and modification operations are usually
smaller than the edited document and they are directly processed by a
transformation engine which adapts them to several representations. In
addition, the possibility to group several selections or modifications
into a single XML document allows us to update multiple views with
different frequencies and fits the requirement of an asynchronous
communication mode like HTTP.

Several clients with different connection speeds can collaborate in a
single editing session. For the moment, specific aspects concerning 
concurrent editing problems have not yet been tackled. We are aware that 
a lot of problems have to be resolved in order to provide a full
featured collaborative environment. Still, we believe that what
we have presented here can be used as a valuable basis for more
elaborated protocols.

XPPML is less expressive than XSLT. Any XPPML specification can be
easily translated into XSLT. XPPML has been designed so that
a dynamic link between the source structure and the transformed one 
could easily be maintained. Doing this for XSLT seems more problematic.
For example, a transformation rule in XSLT can have access to any 
node of the source document, even those localized outside the matched 
subtree. This means that a single modification in the source document can
potentially affect the overall result.

Our system has been successfully tested with the distributed editing
of large Java programs. User-interactions are quickly processed and
reflected to the client. The unique problem of performance we had
to face was only with the initial transmission of the formatted document
which may be very large. We are studying possibilities to 
transmit only the part of the structure needed by the client (the subtree 
visible in the window for example) or to allow several clients to access 
different subtrees of a same document.

For the moment, selections and modifications can only be done on nodes
of the tree. It is sufficient for editing highly structured
structure with few unconstrained text fields, like the kind of
documents we have presented here. However, for editing general XML 
documents, it is necessary to represent selections or
modifications of parts of textual fields. The concept of {\em range}
used in the specification of DOM level 2~\cite{dom2} which represents
a selection by a pair {\em node + offset} will be implemented in the
next version of our system.

In our organization, the amount of software on the side of the
client is kept to a minimum. It is composed of a communication 
layer that sends and receives messages over the network and a layout
engine. All the other components are concentrated in the server. It
is then particularly adapted to situations where clients have sparse
resources.

\section{Acknowledgements}
This work has been done in the framework of Dyade, the Bull-Inria
Research Joint Venture.
%

%
%


\begin{thebibliography}{5}
%

\bibitem {bor88}
P. Borras, D. Cl\'ement, T. Despeyroux, J. Incerpi, G. Kahn, B. Lang and
V. Pascual,
CENTAUR: The system,
Proceeding of the Third Symposium for Software Development
Environments (SDE3),
Boston, December 1988.

\bibitem {cle90}
D. Cl\'ement,
A distributed architecture for programming environments,
Proceedings of the fourth ACM SIGSOFT symposium on Software
development environments,
December 3-5, 1990, Irvine, CA USA.

\bibitem {figue}
B. Conductier, L. Hasco\"et, L. Théry,
Figue's Documentation,\\
see http://www-sop.inria.fr/croap/figue/

\bibitem {css}
CSS2 World Wide Web Consortium.
Cascading Style Sheets, level 2 (CSS2).
W3C Recommendation.
See http://www.w3.org/TR/1998/REC-CSS2-19980512 

\bibitem {dery}
A.M. D\'ery and L. Rideau,
Distributed Architecture for Programming Environment,
INRIA Research Report no 2918, June 1996.

\bibitem {dom2}
DOM-Level-2
W3C (World Wide Web Consortium) DOM Level 2
W3C Recommendation.
See http://www.w3.org/TR/DOM-Level-2.

\bibitem {dsssl}
DSSSL
International Organization for Standardization,
International Electrotechnical Commission.
ISO/IEC 10179:1996.
Document Style Semantics and Specification Language (DSSSL).
International Standard. 

\bibitem {sgml}
ISO 8879:1986, Information processing -- Text and office systems --
Standard Generalized Markup Language (SGML).

\bibitem{shaul}
Israel Z. Ben-Shaul, Gail E. Kaiser and George T. Heineman,
An Architecture for Multi-User Software Development Environments,
Computing Systems, The Journal of the USENIX Association,
6(2):65-103, University of California Press, Spring 1993.

\bibitem {ppml}
Ian Jacobs and Janet Bertot, editors,
Centaur 1.2, chapter The PPML Manual,
Inria Sophia-Antipolis, 1993.

\bibitem {aioli}
L. Th\'ery,
Presentation of A\"ioli,\\
see http://www-sop.inria.fr/lemme/Laurent.Thery/aioli.html.

\bibitem {xpath}
XPath World Wide Web Consortium.
XML Path Language.
W3C Recommendation.
See http://www.w3.org/TR/xpath 

\bibitem {xsl}
XSL World Wide Web Consortium.
Extensible Stylesheet Language (XSL).
W3C Working Draft.
See http://www.w3.org/TR/WD-xsl 

\bibitem {xslt}
XSLT World Wide Web Consortium.
XSL Transformations (XSLT).
W3C Recommendation.
See http://www.w3.org/TR/xslt 

\end{thebibliography}
\end{document}